\documentclass[aps,prb,showpacs,twocolumn]{revtex4}  

\usepackage{graphicx}
\usepackage{color}
\usepackage{amsmath,amssymb}                        %included these packages for symbol support.
\def\gapx{\lower 2pt \hbox{$\buildrel>\over{\scriptstyle{\sim}}$\ }}
\def\lapx{\lower 2pt \hbox{$\buildrel<\over{\scriptstyle{\sim}}$\ }}
\def\he4{$^4$He}
\def\paraH2{{\it p}-H$_2$}
\def\Am2{\AA$^{-2}$}

\begin{document}

\widetext
\title{Adsorption of {\it para}-Hydrogen on Fullerenes}
\author{Joseph D. Turnbull and Massimo Boninsegni} 
\affiliation{Department of Physics, University of Alberta, Edmonton, 
    Alberta, Canada T6G 2J1}
\date{\today}

\begin{abstract}  
Adsorption of  
\textit{para}-hydrogen on the outer surface of a single fullerene
is studied theoretically, by means of ground state Quantum 
Monte Carlo simulations.   We compute energetics and 
radial density profiles of {\it para}-hydrogen for various  coverages on a variety of  small fullerenes.
The equilibrium adsorbed solid monolayer is commensurate 
with the surface of the fullerene; as the chemical potential is increased, a 
discontinuous change is generally observed, to an incommensurate, compressible layer.  
Quantum exchanges of hydrogen molecules are
absent in these systems.
\end{abstract}
  %you might want to verify that these are sane choices.
\maketitle
\section{Introduction}
Low temperature adsorption of highly quantal fluids, such
as helium or {\it para}-hydrogen (\paraH2) on the outer surface
of a fullerene (``buckyball")
can provide insight into physical properties of a quantum many-body system confined 
to spatial regions of nanometer size.  As the diameter of the fullerene is 
increased, the properties of the adsorbate ought to interpolate between 
those of a cluster with a solvated impurity, and those 
of an adsorbed film on an infinite substrate.

In this paper, we consider adsorption of
\paraH2 on a single fullerene C$_l$, with $l$=20, 36, 60 and 80. All of these 
molecules are strong adsorbers, and very nearly spherical.
Background for this study is provided by the wealth of theoretical 
\cite{wagner94,wagner96,gordillo97,Nho1,shi03,boninsegni04} and 
experimental 
\cite {nielsen80,lauter90,wiechert91,vilches92,cheng93b, mistura94, ross98} 
work, spanning over two decades, aimed at investigating the properties of 
adsorbed \paraH2 films on various
substrates. This work is also inspired by recent theoretical 
results on adsorption of helium on buckyballs. 
\cite{Hernandez1, Szybisz1}

A fluid of \paraH2 molecules is an interesting physical system for a number of 
reasons. Because a \paraH2 molecule is half as light as a helium atom, 
zero-point motion can be expected to be quite significant; each molecule is
a spin-zero boson, and therefore it is conceivable that, at low enough temperature,
a \paraH2 fluid might display physical behavior similar to that of fluid 
helium, including superfluidity. \cite{ginzburg72}
Unlike helium, though, bulk {\it p}-H$_2$ solidifies at low temperature 
($T_{\rm c}
\approx$ 14 K); this prevents the observation of phenomena such as 
Bose Condensation and, possibly, superfluidity, which are speculated
to occur in the liquid phase below $T$ $\approx$ 6 K. 
Solidification is due to the depth of the 
attractive well of the potential between two hydrogen molecules, which is
significantly greater than that between two	helium atoms. Several, 
attempts have been made \cite{bretz81,maris86,maris87,schindler96} to 
supercool bulk liquid \paraH2, but the search for superfluidity (in the bulk) has so 
far not met with success.

Confinement, and reduction of dimensionality, are widely regarded as
plausible avenues to the stabilization of a liquid phase of \paraH2 at
temperatures sufficiently low that a superfluid transition may be observed. Indeed,
computer simulations yielded evidence of superfluid behavior in very small
(less than 20 molecules) {\it p}-H$_2$ clusters,\cite{sindzingre91} and claims
have been made of its actual experimental observation. \cite{grebenev00} 
Also, a considerable effort has been devoted, in recent times, to the 
theoretical characterization of
superfluid properties of solvating \paraH2 clusters around linear molecules,
such as OCS. \cite{kwon02,paesani03}

The study of hydrogen adsorption on nanocarbons falls within the same general
research theme, but is also motivated by possible practical applications; an 
important example is hydrogen storage, for fueling purposes. So far, research 
along these lines has mostly focused on nanotubes, 
\cite{dillon97,liu99,wang99,pradhan02,levesque02} but it seems worthwhile to
extend the investigation, possibly providing useful quantitative information
on adsorption on other nanostructures, including fullerenes.

In this work, energetic and structural properties of a layer of \paraH2 
molecules adsorbed on a C$_{l}$ fullerene are investigated 
theoretically, by means of ground state Quantum Monte Carlo (QMC) simulations. 
In order to provide a reasonable, quantitative account  of the 
corrugation of the surface of the fullerene, we explicitly modeled in our study
each individual carbon (C) atom.  For comparison, however,
we have also performed calculations with a simpler model, 
describing fullerenes as smooth spherical surfaces, interacting with 
\paraH2 molecules via an angle-averaged potential. 

Only one adsorbed layer is found to be 
thermodynamically stable on these small nanocarbons.
On a corrugated substrate, a commensurate solid layer is observed at equilibrium; 
as the chemical potential is increased, a discontinuous change 
to an incommensurate solid layer takes place on C$_{20}$, C$_{36}$ and C$_{60}$. 
We could not find, within the statistical uncertainties of our calculation, evidence of an
incommensurate layer on C$_{80}$.  

The difference in compression
between commensurate and incommensurate layers, as measured by the effective 
\paraH2 coverage, is approximately  216\% for C$_{20}$, and decreases to $\sim$ 25\% for C$_{60}$.

Obviously, on a smooth fullerene, there is no distinction between 
commensurate and incommensurate layers.
In the absence of corrugation, energetics of the adsorbed layer are determined 
primarily by the interactions among \paraH2 molecules. The ground state of 
\paraH2 in two dimensions (2D) is a solid, with molecules forming a triangular 
lattice.\cite{gordillo97,boninsegni04b} Our results indicate that \paraH2 
molecules attempt to reproduce  the same triangular arrangement as on an 
infinite plane, even when confined to moving on a spherical surface of radius 
as small as a few \AA.  

Evidence of quantum exchange is absent in these systems, i.e., no evidence 
suggesting possible superfluid behavior is gathered in this work.

The remainder of this manuscript is organized as follows: 
Sec. \ref{model} offers a description of the model used for our system
of interest, including a 
discussion of the potentials used and the justifications for underlying 
assumptions.  Sec. \ref{method} involves a brief discussion of the 
computational technique and specific details of its implementation, in 
addition to details of calibration and optimization.  
The results are presented in Sec. \ref{results}; finally, Sec. 
\ref{conclusions} is a summary of the findings and our concluding remarks.
\section{Model}
\label{model}
We consider a system of $N$ \paraH2 
molecules, in the presence of a single C$_l$ molecule. 
The latter is assumed fixed in
space, owing to its relatively large mass; the center of the molecule is taken
as the origin of a Cartesian coordinate frame.  The $l$ individual C atoms 
are fixed  at positions    $\{{\bf R}_k\}$, $k$=1,2,...,$l$.  
All of the atoms and molecules are regarded as point particles. 
The model quantum many-body Hamiltonian is therefore the following: 
\begin{eqnarray}\label{hm}\nonumber
\hat{H}&=&-\frac{\hbar^{2}}{2m}\sum_{i=1}^{N}\nabla_{i}^{2} + 
\sum_{i<j}V(r_{ij}) + \\ &+ &
\sum_{i=1}^N\sum_{k=1}^l U(|{\bf r}_i-{\bf R}_k|)
\end{eqnarray}
Here, $m$ is the mass of a \paraH2 molecule, $r_{ij}\equiv |{\bf r}_i-{\bf r}_j|$, 
$\{{\bf r}_j\}$ 
(with $j$=1,2,...,$N$) are the positions of the \paraH2 molecules,
$V$ is the potential describing the interaction between any two of them, 
and $U$ represents the interaction of a \paraH2 molecule with a C atom. 
All pair potentials
are assumed to depend only on relative distances.  
The interaction  $V$ is described by the Silvera-Goldman potential 
\cite{Silvera1}, which provides an accurate description of energetic and 
structural properties of condensed \paraH2 at ordinary conditions of 
temperature and pressure.\cite{johnson96,operetto}  
The interaction of a \paraH2 molecule and a C atom is modeled using a 
standard 6-12 Lennard-Jones (LJ) potential, with $\epsilon=32.05$ K and 
$\sigma=3.179$ \AA\ (see, for instance, Ref. \onlinecite{levesque02}).

The model (\ref{hm}) already contains important physical simplifications, such
as the neglect of zero-point motion of C atoms, as well as 
the restrictions to additive pairwise interactions (to the exclusion of,
for example, three-body terms), all taken to be central, and the use of the 
highly simplified LJ interaction. On the other hand,
 (\ref{hm}) is the simplest microscopic model that explicitly takes 
 into account the 
 corrugation of the surface of the buckyball.

A further simplification can be introduced by  replacing the third term in (\ref{hm}) with $\sum_i 
{\tilde U}(r_i)$, where $\tilde U$ is the following, spherically symmetric 
external potential (see Ref. \onlinecite{Hernandez1} for details): 
\begin{eqnarray}\label{sim}
\tilde U(r,{\rm R})&=
&\frac{\epsilon n}{{\rm R}r}\biggl\{ \frac{\sigma^{12}}{5}\left[\frac{1}
    {(r-\rm{R})^{10}} - \frac{1}{(r+{\rm R})^{10}}\right] \nonumber \\ 
    &-& \frac{\sigma^{6}}{2}\left[\frac{1}{(r-\rm {R})^{4}} - 
        \frac{1}{(r+\rm {R})^{4}}\right]\biggr\}
\end{eqnarray}
Here, $n=4\pi \theta$ $a^2$, $a$ being the  of the fullerene and 
$\theta$ being the areal density of C atom on its surface; $\epsilon$ and 
$\sigma$  are the parameters of the LJ potential $V$ introduced for the fully corrugated model.
By using (\ref{sim}), one is describing the fullerene as a smooth spherical 
shell, i.e., corrugation is neglected. 
This approximation substantially simplifies the calculation; it has been 
adopted in recent studies of helium adsorption on buckyballs.\cite{Hernandez1,Szybisz1} 
As mentioned 
above, in this work we have performed calculations based on the full model 
(\ref{hm}), as well as using the effective potential (\ref {sim}); results
obtained in the two ways are compared in Sec.  \ref{results}.

\section{Computational Method}  
\label{method}

Accurate ground state expectation values for quantum many-body systems
described by a Hamiltonian such as (\ref{hm}) can be computed 
by means of QMC simulations. 
In this work, the method utilized is \textit{Variational Path Integral} 
(VPI), which is an extension to zero temperature of the standard, 
Path Integral Monte Carlo (PIMC) method. \cite{Ceperley1}  
VPI (also referred to as Path Integral Ground State, PIGS\cite{Sarsa1}) 
    is a projection technique, which filters out the exact ground state wave 
function out of an initial trial state. It is therefore closely related to
other ground state projection methods, such as Diffusion Monte Carlo (DMC),
but has a few distinct advantages (for a discussion, see, for instance, Ref. 
\onlinecite{Sarsa1}).

The VPI technique works as follows: One samples a large set 
$\left\{X^m\right\}$, $m = 1,2,...,M,$ of discretized many-particle paths 
$X \equiv R_{0}R_{1}...R_{2L}$, where $R_j 
\equiv \textbf{r}_{j1} \textbf{r}_{j2}... \textbf{r}_{jN}$ represents 
the positions of all $N$ particles in the system at the $j$th ``slice".
These paths are sampled based on the probability density
\begin{equation}
{\cal P}(X)\propto \Psi_T(R_0)\Psi_T(R_{2L}) \ \biggl  \{ \prod_{j=0}^{2L-1} G(R_j,R_{j+1},\tau)\biggr \}
\end{equation}
where $\Psi _{T}(R)$ is a trial wave function for the many-body system, 
and $G(R,R',\tau)$ is a  short-time (i.e., $\tau\to 0$)
approximation to the imaginary-time propagator 
$\left\langle R\left|\exp\left[-\tau\hat{H}\right] \right|R'\right\rangle$.  
One can show \cite{Ceperley1} that, for any choice of $\Psi _T$, in the 
limits $L\tau \rightarrow \infty$, $\tau \rightarrow 0$, $R_L$ is sampled 
from the square of the exact ground state wave function $\Phi (R)$.
This allows one to calculate  the ground state expectation value of 
any observable $\hat {\cal O}$ diagonal in the coordinate representation 
as  a simple statistical average
\begin{equation}\label{ave}
\langle\hat{\cal O}\rangle \approx \frac{1}{M}\ \sum_{m=1}^M {\cal O}(R_L^m)
\end{equation}
The ground state energy can be evaluated using the convenient 
``mixed estimator", which yields an unbiased estimate:
\begin{equation}\label{ene}
{\langle \hat{H} \rangle }
    \approx  \sum _{m=1}^{M} \frac{ \hat{H} \Psi _T(R_1^m)}{\Psi _T(R_1^m)} 
\end{equation}
Both expressions (\ref {ave}), (\ref{ene}) can be rendered arbitrarily
accurate by letting $M \rightarrow \infty$.  

The form of $G$ used here is the \textit{primitive approximation}
\begin{equation}\label{primitive}
G(R,R^\prime,\tau) = \rho_F(R,R^\prime,\tau)\  
e^{-\frac{\tau}{2}[\bar U(R)+\bar U(R^\prime)]} + O(\tau^3),
    \end{equation} 
    where $\bar U(R)$ is the total potential energy of the system at the
    configuration $R$, and
\begin{equation}
\rho_{F}(R,R^\prime,\tau) =
    (4\pi\lambda\tau )^{-3N/2}
    \ \prod_{i=1}^N {\rm exp}\biggl [-\frac{({\bf r}_{i}-{\bf r}^\prime_{i})^2}{4\lambda\tau}\biggr ]
    \end{equation}
is the exact propagator for a system of $N$ non-interacting particles.

For a particular choice of $\tau$, since the projection time 
$\lambda=L\tau$ is necessarily finite, simulations must be performed 
for increasing $L$, until one finds that estimates of the observables 
have converged, within statistical errors. 
One must also extrapolate estimates obtained for different $\tau$, 
in order to infer results in the $\tau \rightarrow 0$ limit.  
A more accurate form of $G$ (such as that thoroughly discussed in Ref.  \onlinecite{voth}, accurate to $\tau^{4}$), or a more accurate trial wave 
function $\Psi _T$ allows one 
to obtain convergence with larger $\tau$ and/or shorter $L$, but, at
least in principle, the extrapolated results ought to be independent of the 
choices of $G$ and $\Psi_T$. \cite{note1}
In summary, with sufficient computer time and proper calibration of 
$\tau$ and $L$, one can generate estimates of physical observables that 
are \textit{exact} (errors being only statistical).

Individual simulations involve generating the set $\left\{X^m\right\}$ 
using a random walk through path space.  The same path sampling techniques 
of PIMC are adopted; included are trial moves involving rigid displacements 
of an entire particle path ({\it displace} type moves), and multilevel 
sampling updates (see, for instance, Ref. \onlinecite{Ceperley1}), 
    which are divided into moves of central path portions 
(i.e., not including an endpoint, namely $j$=1 or $j$=2$L$), and moves that 
update endpoint configurations (which involve $\Psi_T$ in the acceptance 
test, as do the \textit{displace} moves).  

The trial wave function utilized is of the Jastrow type:
%\begin{equation}
%\Psi _T(R)=\prod _{i=1}^{N}\exp[-w(r_i)] \prod _{i<j}\exp[-u(r_{ij})]
%\end{equation}
\begin{eqnarray}
\Psi_T({\bf r}_1,{\bf r}_2,...{\bf r}_N)&=& 
\biggl (\prod_{i=1}^N\prod_{k=1}^l e^{-w(|{\bf r}_i-{\bf R}_k|)}\biggr ) \times \nonumber \\  
&\times& \biggl ( \prod_{i<j}^N e^{-u(r_{ij})}\biggr ) 
\end{eqnarray}

which fulfills the symmetry requirement of a many-boson wave function.  In the
case of the spherically averaged model, $l$=1 and $R_{1}$=0.  The
pseudo-potentials $w$ and $u$ were chosen as follows:
\begin{equation}
w(r)=\frac{\alpha}{r^x} \;\; \text{and} \;\; u(r)=\frac{\gamma}{r^5}
\end{equation}
where $x=2$ for the spherically averaged potential, and $x=5$ for 
the corrugated case.  
The values of the parameters $\alpha=80$ \AA$ ^{x}$, $\gamma=750$ 
\AA$^{5}$ and $x$ were obtained empirically, by minimizing the energy 
expectation value computed in separate variational calculations.  

Using the given $\Psi _T$, we observe convergence of the ground state energy 
estimates with a projection time $L\tau$ = 0.250 K$^{-1}$, with $\tau = \tau_\circ = 1/640$ K$^{-1}$.  
Estimates for all quantities quoted in this study were obtained using 
$\tau=\tau_\circ$, even though convergence of the estimates for structural 
quantities can be typically observed with values of $\tau$ significantly 
greater than $\tau_\circ$ (for an equal projection time).

VPI calculations for a range of \paraH2 coverages were carried out for 
each of C$_{20}$, C$_{36}$, C$_{60}$, and C$_{80}$ (we refer here to the
near spherical isomers of each). 
We use an initial configuration of para-hydrogen molecules surrounding the 
fullerene at a distance of approximately twice the C$_l$ radius (i.e.,
making sure that no \paraH2 start off within the buckyball).  
Because of the strongly attractive character of the C$_l$, for a small 
enough number of surrounding hydrogen molecules the system remains spatially
confined (i.e., \paraH2 molecules do not evaporate). Thus, even though periodic boundary 
conditions are used in the 
simulation, they have no effect, so long as a sufficiently large 
simulation cell is used.

The systematic errors of our calculation are attributable to the finite 
projection time $L\tau$ and the finite time step $\tau$.  
Based on comparisons of results obtained from simulations with different 
values of $L$ and $\tau$, we estimate our combined systematic error on the 
total energy per \paraH2 molecule to be of the order of 0.6 K or less.  

\section{Results}  
\label{results}
\begin{figure}
\centerline{\includegraphics[height=2.2in]{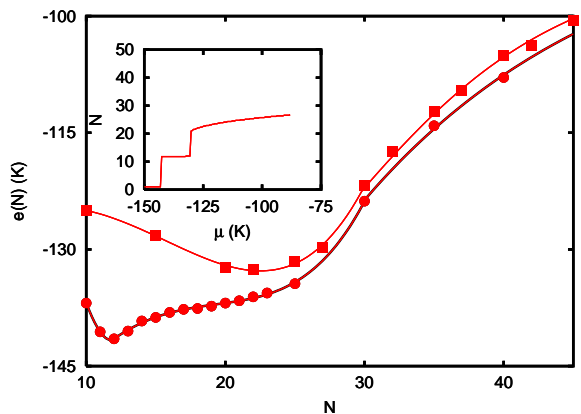}}
\caption{(Color online) Energy per \paraH2 molecule $e(N)$ computed by VPI, as a function 
of the number of molecules adsorbed on a C$_{20}$ fullerene. Filled squares: 
results obtained  with the angularly averaged potential (\ref{sim}). Filled circles:  results obtained by explicitly modeling all carbon atoms in the fullerene. Solid lines are polynomial fits to the VPI data. 
Inset shows the number of particles $N$ plotted as
a function of the chemical potential $\mu$, for the fully corrugated model. The chemical potential of bulk solid \paraH2 is $-88$ K (from Ref. \onlinecite {operetto}), and that is where the $N(\mu)$ curve ends.
}  
\label{energyplot1}
\end{figure}

\begin{figure}
\centerline{\includegraphics[height=2.2in]{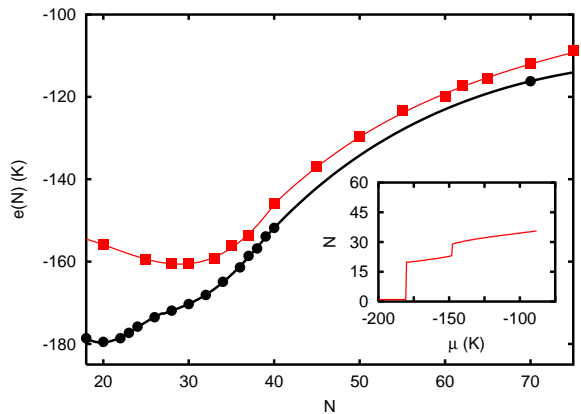}}
\caption{(Color online) Same as Fig. \ref{energyplot1}, but for a C$_{36}$ fullerene. The commensurate layer has $N_C=20$ \paraH2 molecules and is compressible, as evidenced by the finite slope of the curve $N(\mu)$ for 20 $\le N\le$ 28. }  
\label{energyplot2}
\end{figure}

\begin{figure}
\centerline{\includegraphics[height=2.6in,angle=-90]{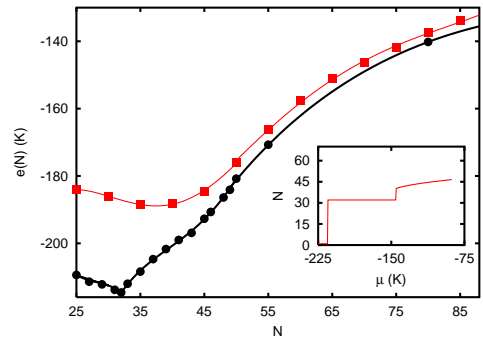}}
\caption{(Color online) Same as Fig. \ref{energyplot1}, but for a C$_{60}$ fullerene. }  
\label{energyplot3}
\end{figure}

\begin{figure}
\centerline{\includegraphics[height=2.2in]{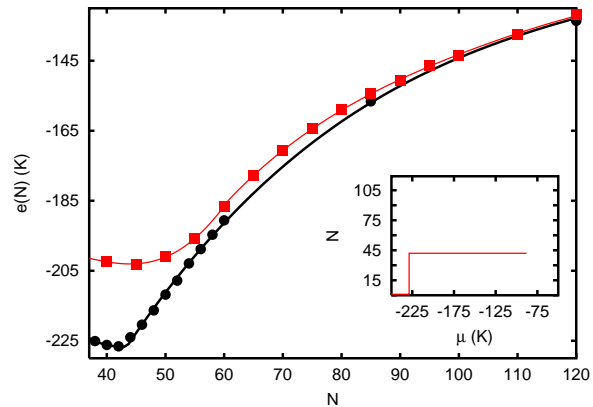}}
\caption{Same as Fig. \ref{energyplot1}, but for a C$_{80}$ fullerene.}  
\label{energyplot4}
\end{figure}

Physical quantities of interest include the ground state energy per \paraH2
molecule $e(N)$ and the radial \paraH2 density $\rho(r)$ about the fullerene, 
as a function of the total number $N$ of molecules. Results for 
$e(N)$ for all fullerenes considered in this study are shown in 
Figs. \ref{energyplot1} through \ref{energyplot4}.

Let us consider first the case of C$_{20}$ (shown in Fig. \ref{energyplot1}), as it allows us to illustrate some of the 
general features seen on other fullerenes as well. 
Filled circles represent the energy estimates yielded by the fully corrugated
model (\ref{hm}), whereas filled squares represent estimates obtained 
using the spherically averaged potential $\tilde U$ described in Sec. \ref{model}. Solid lines are polynomial fits to the numerical data. Also shown in the inset is the number of adsorbed molecules $N$ versus the chemical potential $\mu$ (in K). This is obtained by first fitting the results for $e(N)$, then minimizing the grand canonical potential $\phi(N)=N(e(N)-\mu)$ with respect to $N$, for different values of $\mu$. The chemical potential of bulk solid \paraH2 at zero temperature, computed by Quantum Monte Carlo using the Silvera-Goldman potential,  is $\mu_\circ=-88$ K (from Ref. \onlinecite {operetto}), and that is where all of the $N(\mu)$ curves end.

There are some qualitative similarities, but also important differences (mainly at low $N$) between the results yielded by the two models utilized. Firstly, in the large-$N$ limit they yield essentially the same energy estimates, as corrugation becomes unimportant for thick adsorbed films. This is the case on all fullerenes.

Both $e(N)$ curves display a single 
minimum at a specific number $\bar N$ of \paraH2 molecules.  The minimum corresponds to the mathematical condition $e(\bar N)=\mu(\bar N)$, and physically to the formation of a stable \paraH2 layer. On the corrugated C$_{20}$, a stable layer  occurs for a number $\bar N\equiv N_C=12$,  significantly smaller than that ($\bar N\equiv N_S=22$) obtained on a smooth fullerene.
 Moreover, the energy per molecule $e_C(\bar N_C)$  for the corrugated model, is approximately 10\% lower than that on a smooth substrate  ($e_S(\bar N_S)$), i.e., the corrugated model yields stronger \paraH2 binding.  Qualitatively similar results are observed on all fullerenes; however, the difference between $N_C$ and $N_S$ is seen to decrease with the radius of the fullerene, whereas the difference between the minimum energy values in the two models is $\sim$ 10\% for all fullerenes. 

Corrugation introduces the conceptual distinction between a layer that is {\it commensurate} with the substrate (i.e., the outer surface of the fullerene), and one that is {\it incommensurate} with it. In a commensurate layer, each \paraH2 molecule sits right on top of the center of one of the hexagonal (or, pentagonal) faces of the polyhedron formed by C atoms.  For example, C$_{20}$ has the shape of a 
dodecahedron, i.e. it has 12 pentagonal faces, corresponding to as many adsorption sites for \paraH2 molecules. 
In an incommensurate layer, on the other hand, the arrangement of \paraH2 molecules is mostly determined by their mutual interactions; the fullerene, in this case, merely provides a background attractive potential and a curved geometry. 

On all fullerenes studied here, using the corrugated model,  the equilibrium adsorbed \paraH2 layer is found to be commensurate.  On a corrugated substrate, one normally sees a transition from a commensurate to an incommensurate layer,  as the chemical potential is increased from its equilibrium value.\cite{Nho1} No such transition can be observed on a smooth substrate, for which commensuration is undefined. This is indeed what we {\it generally} observe on the various fullerenes that we have considered. Details, however, differ for the different systems.  

On C$_{20}$, the number of adsorbed \paraH2 molecules remains constant as the chemical potential is increased (i.e., no compression of the commensurate layer is observed), until it jumps discontinuously from $N_C=12$ to $N_I=22$ (see Fig. \ref{energyplot1}). This signals the (abrupt) appearance of an incommensurate layer. The number $N_I$ of molecules in such an incommensurate layer is the same as in the equilibrium layer on a smooth C$_{20}$; this leads us to  surmise that the physics of the incommensurate layer on the corrugated fullerene and that of the equilibrium layer on the smooth fullerene are essentially the same. Thus, the main effect of the neglect of corrugation associated with the spherically averaged potential model, is the absence of the commensurate layer.  We come back to this point later, when discussing density profiles.

The incommensurate layer that forms on a C$_{20}$ fullerene is compressible, i.e., the number of adsorbed molecules is seen to increase monotonically with $\mu$, up to $N\sim28$; no second layer formation is seen before $\mu$ reaches  the value ($\mu_\circ$) corresponding  bulk \paraH2, above which thicker adsorbed films are thermodynamically unstable. In fact, on none of the fullerenes considered here, do we find more than one stable adsorbed layer. 

On C$_{36}$, our results indicate that the commensurate layer is compressible; here too, however, a discontinuous transition occurs to an incommensurate layer, also compressible and again physically similar to the one that forms on a smooth fullerene. 

The same physical behavior  to that observed on C$_{20}$ is seen on C$_{60}$ and C$_{80}$, with an incompressible commensurate layer. On C$_{80}$, though, we fail to observe an incommensurate layer; it should also be noticed that, on this system, it is $N_C=42$ and $N_S=45$, i.e., there is a much smaller relative difference between $N_C$ and $N_S$ than on the other fullerenes.
\begin{table}
  \centering
    \begin{tabular} {|c|c|c|c|c|c|c|c|}
      \hline
C$_l$ & $a$& $N_C$ & $\theta_C$ &  $N_I$ & 
$\theta_I$ &  $N_S$ & $\theta_S$   \\			        
\hline
	C$_{20}$ & 2.00 & 12 & 0.2387 & 22 & 0.4377 & 22 & 0.4377 \\
	\hline
	C$_{36}$ & 2.75 & 20 & 0.2105 & 28 & 0.2946 & 30 & 0.3157 \\
	\hline
	C$_{60}$ & 3.55 & 32 & 0.2021 & 40 & 0.2526 & 40 & 0.2526 \\
	\hline
	C$_{80}$ & 4.11 & 42 & 0.1979 &  - &  -     & 45 & 0.2120 \\				\hline
\end{tabular}
\caption{Effective \paraH2 monolayer coverages for both the corrugated and uncorrugated 
models of the fullerenes. For the model 
including corrugation, $N_C$ and $N_I$ mark the number of \paraH2 molecules  in the adsorbed 
layer at commensuration and incommensuration, respectively, with coverages 
$\theta_C$ and $\theta_I$ (in \Am2). The number  $N_S$ marks the number of \paraH2 molecules  in the
adsorbed layer on a smooth fullerene, $\theta_S$ 
being the corresponding coverage.  The radius $a$ of each fullerene (in \AA) is also given.}
\label{table1}
\end{table}
Our results are summarized in Table \ref{table1}, where the numbers are listed of \paraH2 molecules $N_C$ and $N_I$  in the commensurate and incommensurate layers adsorbed on a corrugated fullerene, as well as $N_S$ for a smooth fullerene. Also shown are the values of  the effective coverage (2D density)
\begin{equation}
\theta=\frac{N}{4\pi {R}^2}
\label{surface}
\end{equation}
where, again, $a$ is the radius of the fullerene.  The effect of the curvature of the substrate can be quantitatively established by comparing these values of $\theta$  to the equilibrium coverage of \paraH2 on a graphite substrate,\cite{Nho1} estimated at 0.070 \Am2. It should be noted that this definition of $\theta$ is used to compare the coverage as a function of {\it the surface area of the substrate}.  A definition of the coverage incorporating the distance at which the \paraH2 sit above the fullerenes is given below (the two definitions would be equivalent for a planar substrate, a distinction only arising with the introduction of non-zero curvature).
\begin{figure}
\centerline{\includegraphics[height=2.7in]{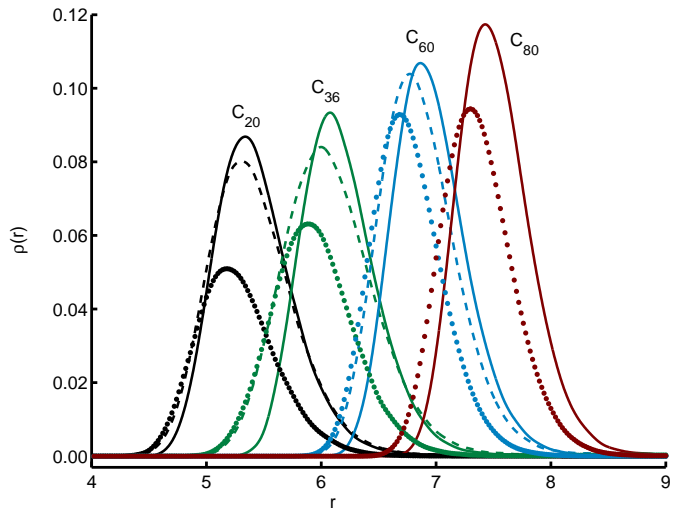}}
\caption{(Color online) Radial density profiles of a \paraH2 layer adsorbed on C$_{l}$.  Solid lines:  profiles obtained for the smooth fullerene model. Dotted lines: profiles of commensurate layers on corrugated model.  Dashed lines: profiles of incommensurate layers on corrugated models. Density is given in \AA$^{-3}$, whereas the distance $r$ from the center of the fullerene is given in \AA.  }  
\label{radialprofiles}
\end{figure}

Structural information about these systems is offered by the radial density profiles $\rho(r)$ of \paraH2 on each of the four C$_l$ for the two models, all shown in Fig. \ref{radialprofiles}. Solid lines denote the radial profiles on a smooth fullerene, while dotted and dashed lines denote radial density profiles
in the corrugated model at commensuration and incommensuration, respectively.   These results are qualitatively the same for each fullerene.   The peak of the profile is shifted away 
from the center of the fullerene, as the system makes a transition from a commensurate to an incommensurate layer, which is not physically unexpected.  No evidence is seen of second layer formation, on any of the fullerenes studied here, for values of the chemical potential for which the adsorbed film is thermodynamically stable (i.e., $\mu\le\mu_\circ$).

On comparing the density profile for the incommensurate layer on a corrugated fullerene (dashed lines) with that for a smooth fullerene (solid lines), one observes a  further shift to the right for the case of  a spherically averaged potential. Perhaps more interestingly, the incommensurate layer on a corrugated fullerene features a greater spatial width, with respect to that on a smooth fullerene (particularly on C$_{20}$ and C$_{36}$). Otherwise, these layers seem physically similar; for example, one can compute  an effective 2D density  $\theta_{eff}$ (different than the coverage $\theta$ discussed above) defined as 
\begin{equation}
\theta_{eff}=\frac{N}{4\pi {a}_{p}^2}
\end{equation}
where $a_{p}$ is the position of the peak of $\rho(r)$. The values obtained for the two layers are very close. In the case of a smooth fullerene, as one goes from C$_{20}$ to C$_{80}$, $\theta_{eff}$ approaches from below the value  0.067 \Am2, namely the equilibrium density of \paraH2 in 2D.\cite{boninsegni04b} This suggests that  the physics of the incommensurate layer (and that on a smooth fullerene) is determined primarily by the interaction among \paraH2 molecules, which attempt to replicate, on a curved surface, the same arrangement as in 2D. 
One further thing to note is that the width of the adsorbed \paraH2 layer on these systems is of the order of 1 \AA, very close to that of an adsorbed monolayer film on graphite.\cite{Nho1}

The computational method adopted here does not allow one to make a direct estimation of the \paraH2 molecule exchange frequency, unlike its finite temperature counterpart (Path Integral Monte Carlo). Nevertheless, from visual inspection of many-particle configurations generated in the Monte Carlo simulation, we observe little or no  overlap of paths associated to different molecules, which is substantial evidence that many-particle permutations are absent  in this system. This is consistent with the high degree of localization that molecules experience, both in the commensurate and in the incommensurate adsorbed layers. 

\section{Conclusions}
\label{conclusions}
Using a numerically exact ground state Quantum Monte Carlo method, we studied  \paraH2 adsorption on the outer surface of near-spherical fullerenes.   We performed calculations based on a simple model, in which all carbon atoms are included explicitly, i.e., corrugation of the surface of the fullerene is captured. 

A  single solid layer of \paraH2 is found to be thermodynamically stable on the fullerenes studied in this work. We find that the equilibrium adsorbed layer is commensurate with the  corrugated surface of the fullerene.  Such a layer is found to be compressible on one of the fullerenes (namely, C$_{36}$), incompressible on the others. On increasing the chemical potential, a discontinuous transition is observed to an incommensurate layer on all  fullerenes, except for the largest one considered here (C$_{80}$).

The basic physics of the incommensurate layer  is driven primarily by the interaction among \paraH2 molecules, which attempt to reproduce the same triangular arrangement as on an infinite plane, even when confined to moving on a spherical surface of radius as small as a few \AA.  Indeed, the incommensurate layer is very similar to that found  using a simpler model of the system, describing the fullerene as a smooth spherical substrate. This simpler model yields results for the energetics and structure of the incommensurate layer in good quantitative agreement with those provided by the fully corrugated model; obviously, however, it is necessary to include corrugation in order to reproduce the commensurate layer. An interesting question that arises is whether a commensurate layer of helium may exist on these molecules; the theoretical studies performed so far have made use of a spherically averaged potential to describe the fullerenes.\cite{Hernandez1}

No evidence of quantum exchanges among \paraH2 molecules has been observed in this work.  Thus, these systems do not appear as not likely candidates for the observation of the elusive superfluid phase of {\it para}-hydrogen.

In conclusion, this study has provided preliminary information on the structure and energetics of hydrogen films adsorbed to the exterior of buckyballs, though much remains to be answered.  The experimental tests of several of the stated predictions could be carried out, for example, by measuring the mass of adsorbed hydrogen on C$_l$ by examination of excitation spectra in a dipole trap. \cite{Han1,Riis1,Gehm1}
\section*{Acknowledgments}
This work was supported in part by the Petroleum Research Fund of the American Chemical Society under research grant 36658-AC5, by the Natural Sciences and Engineering Research council of Canada (NSERC) under research grant G121210893, and by an NSERC PGSA scholarship.


\begin{thebibliography}{99}
\bibitem{note1}
Strictly speaking, there is an additional requirement, namely that $\Psi_T$ 
not be orthogonal to the true ground state wave function. 
This is not a problem for Bose systems, since the ground state wave function 
can always be chosen real and positive, and therefore any positive-definite 
function $\Psi_T$ satisfies the non-orthogonality requirement.
\end{thebibliography}
\end{document}